\documentstyle[aps,preprint]{revtex}
\begin{document}
\draft

\title{The phase diagram of magnetic ladders constructed from a
     composite-spin model}
\author{\"Ors Legeza}
\address{Research Institute for Solid State Physics,
       H-1525 Budapest, P.\ O.\ Box 49, Hungary}
\address{and Technical University of Budapest,
       H-1521 Budapest, Hungary}
\author{G\'abor F\'ath\cite{byline}}
\address{Institute of Theoretical Physics, University of Lausanne,
       CH-1015 Lausanne, Switzerland}
\author{Jen\H{o} S\'olyom}
\address{Research Institute for Solid State Physics,
       H-1525 Budapest, P.\ O.\ Box 49, Hungary}
\date{\today}

\maketitle

\begin{abstract}
White's density   matrix renormalization group ({DMRG}) method has
been applied to an $S= 1/2 + 1/2$ composite-spin model, which can also be
considered as a two-leg ladder model. By appropriate choices of the 
coupling constants this model allows not only to study how the gap is
opened around the gapless integrable models, but also to interpolate
continuously between models with different spin lengths. We have found 
indications for the existence of several different massive phases.
\end{abstract}

\narrowtext
\section{Introduction}

The recent discovery of several families of new mate\-rials,
\cite{johnston,hiroi,cava} such as $({\rm VO})_2{\rm P}_2{\rm O}_7$,
${\rm Sr}_{n-1}{\rm Cu}_{n}{\rm O}_{2n-1}$ ($n=2, 3,\dots$)
and ${\rm La}_{4+4n}{\rm Cu}_{8+2n}{\rm O}_{14+8n}$,
where the spin chains are coupled in a special way to form ladders,
gave a new impetus to the study of the properties of low dimensional
magnetic system. This field was already intensively studied both
experimentally and theoretically in the last decade due to a large
extent to the proposal by Haldane \cite{haldane} that 
antiferromagnetic spin chains described by an isotropic Heisenberg
model develop a gap in  their excitation spectrum for integer spin,
while for half-integer spin  the spectrum is gapless. 

The spin-ladder models, beside their relevance to these materials,
are of special interest for theorists 
\cite{hida,strong,dagotto,watanabe,barnes,hsu,noack,gopalan,tsune,totsu,white95} 
because by appropriate choice of the couplings they can describe both
spin-1/2 and effective spin-1 models, and they are also related to
the models proposed to understand the behavior of the so-called
high-$T_c$ materials. Experimental studies \cite{matter1,azuma}
confirm, that two-leg ladders behave like integer-spin models and
have finite gap, while materials with three-leg ladders have gapless
magnetic excitation spectrum. 

A particular way to construct ladder models is to put a composite
spin on every site of a single chain and to couple the individual
spins in various ways. \cite{solyom1,schulz} This model has already
been studied numerically using the Lanczos algorithm to calculate the
low-lying energy levels. This method could, however, be applied to
relatively short chains  only, and therefore the conclusions were
sometimes contradictory. 

Recently White \cite{white} has proposed a new procedure, the density
matrix renormalization group (DMRG) method that allows to calculate
the energy of low-lying levels and related physical quantities on
much longer chains. This led to a great progress in the application
of the finite-size scaling method and allowed to investigate more
complex systems in a larger parameter space.

A natural extension of the usual Heisenberg model of spin chains for
$S>1/2$ is to include higher order polynomials of the bilinear
exchange term. In the spin-1 case, where biquadratic exchange can be
taken  into account, generically the spectrum remains massive,
\cite{aff} but at some special values of the couplings
\cite{takt,lai} the gap might disappear and the model can be studied
by Bethe's ansatz. The opening of the gap around these critical
integrable points is, however, not quite settled.

The aim of this paper is to extend the earlier calculations on the
composite-spin model to longer chains using the DMRG procedure,
and to clarify the phase diagram of the corresponding ladder model.

The setup of the paper is as follows. In Sec.\ II we give a short
description of the composite-spin and ladder models and their
relationship to the integrable models for appropriate choices of the
parameters. The DMRG method and the numerical procedures are
discussed in Sec.\ III. The results of our numerical calculations are
presented in Sec.\ IV. Finally Sec.\ V contains a brief summary.

\section{The composite-spin and ladder models}

In a  composite-spin model the spin $S_i$  at the lattice site $i$ is
composed of two or more spin operators $\sigma_{i \alpha} \:
(\alpha=1,2, \ldots)$. In the most general case the model contains
on-site and nearest-neighbor interactions among all spin species with
different coupling constants. In this paper we will focus on a model,
where two $ s=1/2$ spin species, from which an $S=1$ spin can be
constructed, are put on every lattice site. 

The model is defined by starting from the most general isotropic
spin-1 model, the bilinear-biquadratic model, which usually is
written in the form 
\begin{equation}     {\cal H}= \/\sum_i \left[ \cos{\theta} 
  \left(\vec S_i \cdot \vec S_{i+1}\right)+
  \sin{\theta} \left( \vec S_i \cdot \vec S_{i+1}\right)^2\right]\,.
\label{eq:biqu}
\end{equation}
In the composite-spin representation, where $\vec S_i$ can be
obtained by adding up the two spin-1/2 species denoted by $\vec
\sigma_i$ and $\vec \tau_i$, $\vec S_i=\vec \sigma_i+\vec \tau_i$,
the Hamiltonian takes the form
\begin{eqnarray}
{\cal H}&=&(\cos{\theta}-{\frac{1}{2}\sin{\theta}})({\cal H}_0+
   {\cal H}_1) + 2\sin{\theta} ({\cal H}_2 +{\cal H}_3) \nonumber \\
&+&\frac{3(N-1)}{4}\sin{\theta}\,,
\label{eq:compbiq}
\end{eqnarray}
where $N$ is the number of lattice sites in the chain, and
\begin{mathletters}
\label{eq:plaquett}
\begin{eqnarray}
{\cal H}_0 & = &\sum_i\left[ \vec\sigma_i \cdot \vec\sigma_{i+1} +
 \vec\tau_i \cdot \vec\tau_{i+1} \right]\,,   \\
{\cal H}_1 & = &\sum_i\left[ \vec\sigma_i \cdot \vec\tau_{i+1} +
  \vec\tau_i \cdot \vec\sigma_{i+1} \right]\,, \\
{\cal H}_2 & = & {\frac{1}{2}}\sum_i\left[ \vec\sigma_i \cdot
\vec\tau_i +\vec\sigma_{i+1} \cdot \vec\tau_{i+1} \right]\,, \\
{\cal H}_3 & = &\sum_i \left[ (\vec\sigma_i \cdot \vec\sigma_{i+1})
  (\vec\tau_i \cdot \vec\tau_{i+1}) \right. \nonumber \\
  & & \left. \phantom{++} + (\vec\sigma_i \cdot \vec\tau_{i+1})
   (\vec\tau_i \cdot \vec\sigma_{i+1}) \right] \,.
\end{eqnarray}
\end{mathletters}
We will generalize Eq.\ (\ref{eq:compbiq}) and consider the model
described by the Hamiltonian
\begin{equation}
{\cal H}= \lambda_0\,{\cal H}_{0}+\lambda_1\,{\cal H}_1+
   \lambda_2\,{\cal H}_2 +\lambda_3\,{\cal H}_3\,,
\label{eq:comp1}
\end{equation}
with arbitrary couplings $\lambda_i$. 

Alternatively, instead of considering $\vec\sigma_i$ and $\vec\tau_i$
as spins sitting on the same site, we can treat them as sitting on
two parallel chains, or on the legs of a ladder, the $\vec\sigma_i$
spins on one leg and the $\vec\tau_i$ spins on the other. As shown in
Fig.\ \ref{fig:ham}, ${\cal H}_0$ couples spins on the same leg only,
the others contain inter-leg couplings. ${\cal H}_2$ is the usual
coupling between spins on the same rung, ${\cal H}_1$ couples spins
on neighboring rungs of the legs, while ${\cal H}_3$ describes
four-spin couplings on a plaquette. 

Usually the ladder models are constructed to include ${\cal H}_0$ and
${\cal H}_2$ only. When a strong ferromagnetic coupling is applied
across the rungs $(\lambda_2 \to -\infty)$, the two spins form a
triplet  and the properties of the $S=1$ Heisenberg chain are
recovered. \cite{hida,watanabe} In another approach Barnes {\em et
al.} \cite{barnes} allowed for strong antiferromagnetic interchain
couplings and treated  ${\cal H}_0$ as perturbation. They have shown
the existence of a spin gap  for any finite interchain coupling.

In the composite-spin model, on the other hand, ${\cal H}_1$ is also 
necessarily included. A special feature of the model is that for 
arbitrary values of $\lambda_2$ and $\lambda_3$ the model is invariant 
under the interchange of $\lambda_0$ and $\lambda_1$, i.e.,  the energy 
levels of the full Hamiltonian satisfy
\begin{equation}
     E(\lambda_0, \lambda_1, \lambda_2, \lambda_3) =
          E(\lambda_1, \lambda_0, \lambda_2, \lambda_3) \,.
\label{eq:lam_0-lam_1}
\end{equation}
This can be shown by interchanging the $\vec\sigma_i$ and
$\vec\tau_i$ spins on every second site. This relationship allows to
connect the weak- and strong-coupling limits of the model by a
duality transformation. To show this let us denote by $\varepsilon$
the energies of the Hamiltonian in which the coupling strength of
${\cal H}_0$ is chosen to be unity. 
\begin{equation}
     E(\lambda_0, \lambda_1, \lambda_2, \lambda_3) = \lambda_0 \,
    \varepsilon({\lambda_1 \over \lambda_0},{\lambda_2 \over \lambda_0},
   {\lambda_3 \over \lambda_0}) \,.
\end{equation}
From Eq.\ (\ref{eq:lam_0-lam_1}) it follows that
\begin{equation}
     \lambda_0 \,\varepsilon({\lambda_1 \over \lambda_0},{\lambda_2 
    \over \lambda_0}, {\lambda_3 \over \lambda_0}) = \lambda_1 \,
    \varepsilon({\lambda_0 \over \lambda_1},
    {\lambda_2 \over \lambda_1}, {\lambda_3 \over \lambda_1}) \,.
\end{equation}
Introducing the couplings $\widetilde{\lambda}_i = \lambda_i / \lambda_0$
$(i=1,2,3)$, we get
\begin{equation}
     \varepsilon(\widetilde{\lambda}_1 , \widetilde{\lambda}_2 ,
  \widetilde{\lambda}_3 ) = \widetilde{\lambda}_1 \,
  \varepsilon({1 \over \widetilde{\lambda}_1}, {\widetilde{\lambda}_2 
   \over  \widetilde{\lambda}_1}, {\widetilde{\lambda}_3 \over
     \widetilde{\lambda}_1}) \,.
\end{equation}
In what follows we will always work with the Hamiltonian in which
$\lambda_0 =1$ and will drop the tilde over the couplings.

For $\lambda_2=\lambda_3=0$ this relationship reduces to a usual
duality relationship, which connects the energies in the $0 < 
\lambda_1 \leq 1$ region to those in $1\leq\lambda_1<\infty$,
\begin{equation}
    \varepsilon(\lambda_1)=\lambda_1 \,\varepsilon(1/\lambda_1) \,.
\label{eq:self}
\end{equation}

When $\lambda_1 < 0$, the above deduced duality relationship connects
the lowest-lying levels of one region to the highest-lying levels in
the other region. Since in the numerical calculations a few low-lying
levels can only be calculated with sufficient precision, a more
useful relation can be derived in this case by comparing the energies
of ${\cal H}$ defined by Eq.\ (\ref{eq:comp1}) with $\lambda_0 = 1$
to that of ${\cal H}' = - {\cal H}$. Denoting by
$\varepsilon'(\lambda_1, \lambda_2, \lambda_3)$ the energies of this
model, 
\begin{equation}
     \varepsilon'(\lambda_1, \lambda_2, \lambda_3 ) = - \lambda_1
    \varepsilon({1 \over \lambda_1}, { \lambda_2 \over \lambda_1}, 
    {\lambda_3 \over \lambda_1}) \,.
\label{eq:self1}
\end{equation}
We used these relations to check the accuracy of the numerical 
calculations.

According to Eq.\ (\ref{eq:compbiq}) the spin-1 model can be
reproduced when ${\cal H}_0$ and ${\cal H}_1$ are taken with equal
coupling strength, $\lambda_0 = \lambda_1$. When this coupling is
antiferromagnetic, it gives rise to a valence-bond configuration
\cite{vbs} of the neighboring spins, which is a characteristic
feature of the Haldane phase. When no direct coupling exists between
the $\vec\sigma_i$ and $\vec\tau_i$ spins, nothing ensures a priori
that these spins appear in the symmetric $S=1$ configuration only.
When, however, $\lambda_2 $ and $\lambda_3$ are small compared to
$\lambda_0 = \lambda_1$, the level structure of the true spin-1 model
and that of the composite-spin model is such that their low-lying
parts coincide, thus their behavior is similar. Therefore the phase
diagram of our ladder model will contain phases characteristic to the
bilinear-biquadratic model, but also new phases may appear.  

The normalization $\lambda_0 =1$ is equivalent to considering the 
region $-\pi/2 < \theta < \pi/2 $ only. The bilinear-biquadratic
model has a rich phase structure in that region. It has been
thoroughly investigated both analytically \cite{aff,parkinson} by
using, e.g., the mapping to the Wess-Zumino-Witten model, or to the
9-state Potts model  and numerically. \cite{kung,solyom,tak,gfath}
The region $\pi /2 < \theta < 3\pi / 2$ is somewhat less interesting,
since the model is ferromagnetic for $\pi / 2 < \theta < 5\pi / 4$.
Beyond that a new phase might appear, as predicted by Chubukov,
\cite{chubukov} although so far its existence has not been confirmed
by numerical calculations. \cite{gfath95} This problem is, however,
outside the scope of this paper.

$\theta_{\rm TB}= - \pi / 4$ corresponds to the
Takhtajan-Babujian \cite{takt} integrable model with gapless
excitation spec\-trum. This point is the critical point of a
second-order Ising-type phase transition with a gap opening linearly
around the transition point. In a chain with periodic boundary
condition (PBC) the ground state is doubly degenerate for $\theta<-
\pi / 4$ producing a dimerized phase, while for $- \pi/4 < \theta <
\pi / 4$ the Haldane phase appears with a non-degenerate  singlet
ground state. This latter region includes the isotropic Heisenberg
point at $\theta=0$ and the exact nearest neighbor valence-bond state
(VBS)  \cite{vbs} at $\theta_{\rm VBS}=\arctan (1 /3)$.

$\theta_{\rm LS}= \pi /4$ is  another integrable point related
to the Lai-Sutherland \cite{lai} model, where the gap vanishes again.
For $\theta > \pi / 4 $ a trimerized massless phase appears. 
\cite{tak,gfath,itoi} In the composite-spin representation these
points  lie on the line $\lambda_1 =1$, at $\lambda_2 = \lambda_3 =
-4/3$ for the  Takhtajan-Babujian, at $\lambda_2 = \lambda_3 = 0$ for
the Heisenberg,  at $\lambda_2 = \lambda_3 = 4/5$ for the VBS and at
$\lambda_2 = \lambda_3 = 4 $ for the Lai-Sutherland points. 

Furthermore, the composite-spin Hamiltonian can be transformed into a
nonlinear $\sigma$ model and the results obtained in field theory can
be related to our model. At $\lambda_1=\lambda_2=\lambda_3 =0$, when
the model reduces to two decoupled spin-1/2 chains, the field theory
possesses a gapless excitation spectrum because the topological angle
is an odd multiple of $\pi$. According to the field-theoretical
calculations \cite{aff} the perturbations around this multicritical
point are generically relevant and give rise to the opening of an
energy gap, except for special points or lines. The
Takhtajan-Babujian and Lai-Sutherland models belong to such
exceptions. Consequently, it is expected that by varying  the
$\lambda_i$ parameters the gap vanishes only at the critical  
points and  on the phase boundaries between the various phases.

In order to simplify the calculations, first we study the effect of
${\cal H}_1$, ${\cal H}_2$ and ${\cal H}_3$ separately for fixed 
$\lambda_0=1$. Then the calculation will be extended to a
two-parameter plane, $(\lambda_1,\lambda_2)$ by choosing $\lambda_3 =
\lambda_2$. In order to determine the phase diagram we examine the
low-lying energy levels along different paths connecting those points
in the parameter space,  where the model is integrable. The
ground-state configurations will also be  investigated by calculating
the local energy of a bond, the two-point  correlation functions and
a short-range order parameter.  

\section{Numerical procedures}

We have performed numerical calculations by applying the DMRG method 
\cite{white} on the model defined by the Hamiltonian in Eq.\ 
(\ref{eq:plaquett})-(\ref{eq:comp1}). This is a real-space
renormalization method where the lattice is built up gradually to the
desired length,  systematically truncating in the meantime the
Hilbert space by keeping  only the most probable states.

Since the DMRG method works best for systems with free ends, we will 
consider our composite-spin model with open boundary condition (OBC).
An  unfavorable consequence of OBC for the spin-1 model and
consequently for  the ladder models as well is that the degeneracies
in the spectrum may be different from that obtained for a closed
ring. In the nearest-neighbor  valence-bond configuration, e.g., free
$s=1/2$ spins remain at the ends of the chain, giving rise to a
fourfold degenerate ground state. In the dimerized state, on the
other hand, the twofold degeneracy of the ground state is lifted if
the number of sites is an even number. This makes the  analysis of
the spectrum more difficult.  

Moreover, in the case of OBC the total momentum is not a good
quantum number. Therefore, only the total spin $S_T$ and its
projection  to the quantization axis, $S_T^z$ can be used to classify
the energy levels.  Since in the isotropic case the $SU(2)$ symmetry
is satisfied, the  spectrum was analysed by calculating a few
low-lying levels of the  different $S_{T}^z$ sectors. $S_T$ was
determined from the degeneracy  of the levels.

Because our aim is to determine the overall behavior of the energy
spectrum and to identify from it the possible phases, in most of the
calculations we have used the less accurate version of DMRG, the
so-called {\em infinite-lattice method}. In this algorithm the
lattice is  built up by adding two lattice sites in each step. The
results obtained  for chains with $N=4,8,16,32,48,64$ sites were used
in a finite-size scaling  procedure to extrapolate to infinite
lattice. On the other hand, close to the critical points and in those
regions where the gap is small, the  {\em finite-lattice method} was
applied with two or three  iteration cycles to determine the energies
more precisely. 

In some cases, before doing the DMRG calculation on long chains, the
energy spectrum was determined by exact diagonalization on short
chains with $N=4,6,8$. This gave an idea of the sequence of the
levels for different  choices of the couplings and allowed to
determine which states have to be  targeted in the DMRG procedure.

Considering our limited computational resources, we had to restrict
rather drastically the number of states, $M$ to be kept in the DMRG
algorithm. A number, that can be used to characterize the numerical
accuracy  of the DMRG method is the discarded density-matrix weight
(truncation error).  \cite{white} In the calculation of the
ground-state energy for ladders with $N=2\times 64$ ($N=2\times 32$)
sites it was largest close to the critical  points with values of the
order $10^{-5}-10^{-6}$ ($10^{-7}-10^{-8}$), while around the VBS
point, where the gap is large, it was as small as $10^{-10}-10^{-11}$
($10^{-11}-10^{-12}$). For the excited states the  truncation error
was worse by one order of magnitude. 

The real error in the DMRG procedure can, however, be much worse.
\cite{legeza} It was estimated by comparing the energies obtained for
the ladder model using the DMRG procedure to those of the spin-1
bilinear-biquadratic  model or the spin-1/2 chain, where much better
accuracy can be achieved. For chains with $N=2\times 32$ sites an
agreement up to three or four  decimal places has been achieved after
the second iteration cycle of the  finite-lattice method, if $M$ was
chosen to be $M=64$, while for chains  with $N=2\times 64$ sites the
same accuracy required to retain $M=76$ or  $84$ states. At the VBS
point the infinite-lattice algorithm without the  iteration cycles
provided almost the same accuracy.  

In the extrapolation procedure the $N\to \infty$ limit of the gap was
determined by fitting a form $\Delta E(N) = \Delta + a/N $ or $\Delta
E(N) =  \Delta + b/N^2 $ to the energy differences measured between
the energy  levels of the various $S_T$ sectors. In principle the
first has to be used  when the gap vanishes, $\Delta = 0$, while the
second gives  the correct asymptotic behavior when the gap is finite.
\cite{sorensen}  When the gap is finite but the chain length is not
long enough to observe the correct parabolic behavior, the linear fit
gives a lower-bound estimate.  \cite{schollwock} We have used the
parabolic fit, whenever deviation from  the linear dependence on
$1/N$ was observed.  

\section{Numerical results}

In this section we present our numerical results for the low-lying
levels of the model for various choices of the couplings. For a
finite ladder  the ground state is always a spin singlet. Above this
level there are two  low-lying triplet excitations, which at
$\lambda_1 = \lambda_2 = \lambda_3  = 0$ have equal energy. At this
point, they correspond to independent  excitations on the two
decoupled legs. These levels are followed by several
$S_T=0,1,2,\ldots ,$ excitations. The energy differences between all
these  levels scale as $1/N$, indicating a gapless excitation
spectrum.  

For finite values of $\lambda_1, \lambda_2$ and $\lambda_3$ the
degeneracy of  the two low-lying triplet levels is in general lifted.
This is shown  along the $\lambda_1$ axis in Fig.\ \ref{fig:eigv}.
In a finite chain with $N$ sites the energy difference between the
lowest  triplet and the ground-state singlet levels will be denoted
by  $\Delta E_{10}(N)$. When this quantity scales to a finite value,
the  ground state is a non-degenerate singlet with a finite gap.
When, however,  it scales to zero, we have to consider the next
level. This is either the  second triplet, or it may have been
crossed by the lowest $S_T = 2$ level.  The energy difference between
this quintuplet and the ground-state singlet  levels will be denoted
by $\Delta E_{20}(N)$, while $\Delta E_{21}(N)$ denotes  the energy
difference between this quintuplet and the lowest triplet levels.  

\subsection{Gap opening due to ${\cal H}_1$}

Let us consider first the effect of ${\cal H}_1$, since earlier
calculations  \cite{solyom1} by exact diagonalization on relatively
short chains could not  determine satisfactorily the way the gap is
opened. For $\lambda_1 >0$ the  energy of the lowest triplet states
and that of the singlet ground state come exponentially close to each
other, and become degenerate in the thermodynamic limit
($N\to\infty$). This gives the well known fourfold degeneracy of the
VBS-like state in a finite chain with OPB. 

Therefore the relevant energy gap can be obtained most easily by
measuring  the energy difference $\Delta E_{21}(N)$ between the
lowest-lying levels of  the $S_T=2$ and $S_T=1$ sectors. This energy
difference as a function of  $\lambda_1$ is shown in Fig.\
\ref{fig:l1} for $\lambda_1 > 0$. The lines  connecting the values
calculated at several distinct points are only guides  to the eye.
The inset in the figure shows in more detail the behavior of  the gap
close to the integrable point. 

In the extrapolation to $N\to\infty$ a $1/N$ fit was applied close to
the  critical point, because the chain lengths were not long enough
to observe  the parabolic behavior. As an indication of the accuracy
of our calculation  we mention, that at $\lambda_1=1$, where the
Haldane gap should be recovered, we obtained $\Delta_{21}=0.41(1)$,
in reasonable agreement to three decimal  places with its best
estimate. \cite{sorensen,white3} The ground-state  energy divided by
the chain length converges to $E_{0}/N=1.401484(0)$, in  agreement to
six decimal places with earlier calculations. \cite{white3}  

As a further check we have also performed calculations for large
values of $\lambda_1$ and used the self-duality relationship in Eq.\
(\ref{eq:self}) to obtain the gap for weak couplings. The results
obtained in this way  are shown in the inset. From this we could also
conclude that  our finite-size calculations are correct to at least 4
decimal places. Even though the accuracy of our calculations is
limited, we have found a  linearly opening energy gap for $\lambda_1
>0$. This is the first calculation where this could be demonstrated
numerically. 

For negative values of $\lambda_1$ the coupling between the chains
is ferromagnetic, while the `on-chain' coupling remains
antiferromagnetic. The low-lying part of the spectrum is essentially
different from that of the $\lambda_1>0$ regime. The degeneracy of the
two $S_T=1$ triplet levels is again lifted, but even the lowest of
them will  remain separated from the singlet ground state. So the
energy difference  to be studied is $\Delta E_{10}(N)$. This quantity
is shown on the left  hand side of Fig.\ \ref{fig:l1} for $\lambda_1
< 0$. As it can be seen, the  ground state is a non-degenerate
singlet with a small, but finite gap  to the lowest lying
excitations.   

Since the gap is small, the value of $\Delta_{10}$ has been
calculated using the finite-lattice method. This procedure with three
iteration cycles and a $1/N$ fit gave at $\lambda_1 = -1$, e.g.,
$\Delta_{10} = 0.11(4)$ as a lower-bound estimate for the
singlet-triplet gap. Furthermore, the lowest quintuplet level has
been found to become degenerate  with the lowest triplet state in the
thermodynamic limit. Thus a gap separates  the ground state from the
continuum of excitations. 

For small values of $\lambda_1$ the accuracy of the calculations has
been  checked using the relationship in Eq.\ (\ref{eq:self1}). The
gap is found to  open linearly. Thus ${\cal H}_1$ is relevant for
both signs of $\lambda_1$. 

On the other hand, for large negative values, in the region
$\lambda_1 < -10$, we have found the vanishing of both $\Delta_{10}$
and $\Delta_{20}$.  This indicates that a new phase with gapless
spectrum may appear there.  However, due to the smallness of the gap
around $\lambda_1 = -1$ we were  unable to locate the point where the
transition occurs.

\subsection{The effect of ${\cal H}_2$ and ${\cal H}_3$}

The effect of the inter-leg coupling ${\cal H}_2$ on the decoupled
chains described by ${\cal H}_0$ for $\lambda_1 = \lambda_3 = 0$ has
been  considered by several authors. \cite{hida,watanabe,hsu} Hida
found that  the gap is finite for $ \lambda_2 < \lambda_{2c}=-0.6$,
while Dagotto  {\em et al.}\/ \cite{dagotto} showed some evidence
that the critical  coupling is closer to zero, perhaps $\lambda_{2c}
\approx -0.4$. On the  other hand Watanabe {\em et al}.\
\cite{watanabe} and Hsu and Angl\`es  d'Auriac \cite{hsu} argued that
the interchain coupling defined by  ${\cal H}_2$ is always relevant
and $\lambda_{2c}=0$. 

We repeated some of these calculations to confirm that this coupling
is in fact relevant for both signs of the coupling $\lambda_2$. For
$\lambda_2 > 0$  the ground state is a non-degenerate singlet even in
the thermodynamic limit  with a rather large gap. At $\lambda_2 =
4/3$, e.g., a singlet-triplet gap  with $\Delta_{10}=0.72(9)$, and a
singlet-quintuplet gap with  $\Delta_{20}=1.43(4)$ has been found. 

For $\lambda_2 < 0$ the spectrum is similar to that for $\lambda_1 >
0$,  where the ground state becomes fourfold degenerate in the
thermodynamic limit. The gap, however, is rather small. At
$\lambda_2=-4/3$ the  calculation with the finite-lattice method and
a $1/N$ fit gave  $\Delta_{21} = 0.11(3)$, in agreement with previous
results. \cite{watanabe} 
 
The plaquette coupling ${\cal H}_3$ also turned out to be relevant
for both signs of the coupling, but in some sense its effect is
opposite to that of ${\cal H}_2$. A fourfold degenerate ground state
is obtained with a finite quintuplet-singlet gap for $\lambda_3 > 0$.
At  $\lambda_3=4/3$, e.g.,  $\Delta_{20}=0.28(5)$ has been found. For
$\lambda_3 < 0$ the ground state remains singlet. At $\lambda_3=-4/3$
the triplet-singlet gap is $\Delta_{10}=0.26(8)$. 

Because of this opposing effect of these couplings, next we
considered the competition of ${\cal H}_2$ and ${\cal H}_3$ by
choosing $\lambda_2 =  \lambda_3$. For negative values of $\lambda_2$
the low-lying part of the  energy spectrum resembles very much that
found above for small negative  $\lambda_1$. A finite gap develops
between the singlet and triplet levels as  shown in Fig.\
\ref{fig:l2}. At $\lambda_2=-4/3$ the finite-lattice algorithm  and a
$1/N$ fit gave $\Delta_{10}=0.17(5)$. Close to the critical point the
extrapolated lower-bound values of the gap are even smaller,
therefore  the available chain lengths are still not long enough to
determine the  character of the opening of the gap.  

On the other hand, for $\lambda_2=\lambda_3 > 0$ the analysis of the
low-lying  energy spectrum has shown that both $\Delta_{10}$ and
$\Delta_{20}$ scale  to zero, i.e.\ the spectrum is gapless. Since
the system remains critical in an extended region for $\lambda_2 =
\lambda_3 >0$, a  Kosterlitz-Thouless-like transition may occur at
$\lambda_1 =\lambda_2 = \lambda_3 =0$ to the massive phase at
$\lambda_2 = \lambda_3 < 0$. This could explain the slow opening of
the gap in that region. 

These results are only partially consistent with the field
theoretical predictions, which states that the spin-1/2 integrable
point is unstable  against all perturbations. Our calculations
indicate that although both ${\cal H}_2$ and ${\cal H}_3$ are
relevant operators in the ladder model for both signs of the
couplings $\lambda_2$ and $\lambda_3$, the model remains critical at
least along the $\lambda_2 = \lambda_3 >0$ half line. 

To summarize our findings, three types of the spectrum have been
found. An asymptotically fourfold degenerate ground state,
charateristic of a VBS-like state is obtained for $\lambda_1 > 0$,
$\lambda_2 = \lambda_3 =0$, for $\lambda_2 < 0$, $\lambda_1 =
\lambda_3 =0$ and for  $\lambda_3 > 0$, $\lambda_1 = \lambda_2 =0$. A
truely non-degenerate singlet ground state is found for $\lambda_1 <
0$, $\lambda_2 = \lambda_3 =0$, for $\lambda_2 > 0$, $\lambda_1 =
\lambda_3 =0$, for  $\lambda_3 < 0$, $\lambda_1 = \lambda_2 =0$ and
also for  $\lambda_2 = \lambda_3 < 0$, $\lambda_1 =0$. Finally the
spectrum is gapless for $\lambda_2 = \lambda_3 > 0$, $\lambda_1 =0$.

\subsection{Competition of ${\cal H}_1$, ${\cal H}_2$ and ${\cal
H}_3$ } 

After having determined the spectrum along the $\lambda_i$ axes and
the  $\lambda_2=\lambda_3$ line, next we consider the phase diagram
in the parameter space spanned by $\lambda_i$. To simplify the
calculations we restrict ourselves in the remaining part of the paper
to a two-parameter  plane by choosing $\lambda_2 = \lambda_3$. We
will study especially  the neighborhood of the $\lambda_1=1$ line,
which corresponds to the  bilinear-biquadratic Hamiltonian.  

In the Takhtajan-Babujian point of the bilinear-bi\-quadratic
model, which corresponds to $\lambda_1=1$, $\lambda_2= \lambda_3
=-4/3$ in  our model, the spectrum is gapless in the thermodynamic
limit. Both the singlet-triplet ($\Delta E_{10}(N)$) and the
triplet-quintuplet  ($\Delta E_{21}(N)$) energy differences vanish as
$1/N$. They behave, however, quite differently as we move away from
the critical point. This is shown in Figs.\ \ref{fig:l1_1a} and
\ref{fig:l1_1}. 

For $\lambda_2>-4/3$ the lowest triplet level becomes asymptotically
degenerate with the ground-state singlet, so the relevant gap is
between the  triplet and quintuplet levels ($\Delta_{21}$). This
level structure is the  same as along the $\lambda_1 > 0$ line, thus
a VBS-like state is obtained in this part of the phase space. The
opening of the gap is very slow, even  our longest chains are too
short to obtain a reliable estimate of the gap  close to the critical
point. The error of the extrapolated value close to the critical
point is indicated in Fig.\ \ref{fig:l1_1a} by the size of the  
symbol.   
   
For $\lambda_2<-4/3$ a finite gap develops between the ground-state
singlet  and the lowes triplet levels, in the same way as for
$\lambda_2 =  \lambda_3 <0$. A lower-bound estimate of the gap
obtained by a $1/N$ fit is shown in Fig.\ \ref{fig:l1_1}. The results
are in agreement  with a linearly opening extra\-polated gap
$\Delta_{10}$.   

It is known, however, that the $\lambda_2<-4/3$ region corresponds to
a  dimerized phase, where instead of a singlet ground state a doubly
degenerate  ground state would have to be found. Therefore, we have
considered several  higher lying levels to search for the other
singlet level that would become degenerate with the ground state. We
have not found any such level. This can be understood by recalling
that in the case of OPB the dimerized phase gives a truely twofold
degenerate ground state only if the number of sites is odd, i.e., the
number of bonds is even. For chains with even number of sites the
energy of the dimerized state depends on  whether the bonds at the
ends are strong or weak. Inspection of the low-lying levels is
therefore not sufficient to distinguish a real non-degenerate
singlet ground state from a dimerized state. We will return to this
problem later, when the dimer order parameter will be discussed.

One of the two kinds of behavior found for $\lambda_2 = \lambda_3 >
-4/3$ and $\lambda_2 = \lambda_3 < -4/3$ appears whenever one moves
away from the  Takhtajan-Babujian integrable point in any direction
in the $(\lambda_1, \lambda_2=\lambda_3)$ plane. Varying, e.g.,
$\lambda_1$ around  $\lambda_1=1$ at $\lambda_2=\lambda_3=-4/3$, a
non-degenerate singlet ground state is found for $\lambda_1<1$, while
for $\lambda_1>1$ the singlet-triplet gap disappears forming a
fourfold degenerate ground state.  This result is in agreement with
the assumption that the Takhtajan-Babujian point is generically
unstable against perturbations, except along the phase boundaries.

Our calculation is, however, not accurate enough to locate this
critical line  in the phase space that separates the two kinds of
behavior. In the  schematic phase diagram shown in Fig.\
\ref{fig:phase} the boundaries are  therefore indicated by wavy
lines. That the phase boundary between the VBS-like Haldane phase and
the supposedly dimerized phase connects the $\lambda_1 = \lambda_2 =
\lambda_3 =0$ and the Takhtajan-Babujian points has been confirmed by
looking at the energy spectrum of our model along the trajectory
parametrized by $\lambda_1=1-\lambda$, $\lambda_2 = \lambda_3 = -4
\lambda /3$, where $\lambda$ varies between 0 and 1. For small values
of $\lambda$  a fourfold degenerate ground state is recovered, while
for $\lambda$ close to unity a non-degenerate singlet ground state is
found. 

As mentioned before, this singlet ground state was found along the
$\lambda_1 < 0$, $\lambda_2 = \lambda_3 =0$ line as well. Whether
this axis also belong to the dimerized phase or not, cannot be
determined from the  low-lying spectrum alone. We will return to this
problem later. 

Extending the calculations to $\lambda_2 = \lambda_3 > 0 $ on the
line $\lambda_1=1$ the VBS-like state survives in a finite range,
including the point $\lambda_2= \lambda_3 = 4 /5 $, where the exact
nearest-neighbor valence bond state is recovered. At this point the
lowest singlet and  triplet levels are degenerate for any finite
chain lengths. The gap reaches  its maximum value with
$\Delta_{21}=0.8404(7)$, which after including the  appropriate
scaling factors due to our normalization to $\lambda_0 =1$,   agrees
to 5 digits with the known result. \cite{gfath3}  

For larger values of $\lambda_2$ along the $\lambda_1 = 1$ line the 
Lai-Sutherland model and the trimerized phase of the
bilinear-biquadratic  model cannot be reached by this ladder model.
The extra levels introduced  by the composite-spin representation
will be low lying and will lead to a  gapless new phase, as has
already been found for $\lambda_1 =0$.   This massless phase has been
found for large positive $\lambda_2 = \lambda_3$ values also for
$\lambda_1 < 0$. Thus it is stable in an extended range of the
couplings. 

\subsection{Dimer order}

Even though the structure of the low-lying part of the energy
spectrum  indicates the existence of various phases, its knowledge,
as discussed above,  is not sufficient to clarify unambiguously the
character of the ground state. We have, therefore, calculated several
quantities in the ground state, like the local magnetization $\langle
\vec S_i\rangle \equiv \langle\vec  \sigma_{i}\rangle+\langle\vec
\tau_i \rangle$, the two-point correlation  function $\langle \vec
S_i \vec S_j \rangle \equiv \langle  \left( \vec
\sigma_i+\vec\tau_i\right) \left( \vec \sigma_j+\vec\tau_j\right)
\rangle$ and the local energy $E_{\rm loc} \equiv \langle {\cal
H}(i,i+1)\rangle$, where ${\cal H}(i,i+1)$ contains the couplings 
between spins on sites $i$ and $i+1$.

The correlation function falls off exponentially both in the Haldane 
and the dimerized phases, turning to power-law like on the phase
boundary, but the chains are still too short to distinguish clearly
between these two possibilities. A better procedure could be to look
at the local magnetization in the lowest triplet state.  Due to the
free end spins of the exact nearest-neighbor valence-bond
configuration the local magnetization is finite close to the chain
ends  in the VBS-like Haldane phase. Approaching the phase boundary,
the extra  spin becomes less and less localized and it spreads out
homogeneously in the dimerized and gapless phases.  

The most useful procedure is, however, to consider the so-called 
short-range dimer order parameter. It can be defined by taking the
difference of the local energy on neighboring bonds in the middle of
the ladder,
\begin{equation}
  R_{\rm srdo} = \frac{\langle {\cal H}(i,i+1) \rangle -
  \langle  {\cal H}(i+1,i+2) \rangle}
{\frac{1}{2}\left[ \langle {\cal H}(i,i+1) \rangle+
\langle {\cal H}(i+1,i+2) \rangle\right] } \,.
\end{equation}
It is expected to have different behavior if the ground state is unique,
fourfold degenerate, or twofold degenerate as in the dimerized state 
with spontaneously broken translational symmetry.

A simpler quantity can be defined by taking the bilinear part of the 
coupling only 
\begin{equation}
  S_{\rm srdo} = \frac{\langle \vec S_{i} \cdot \vec S_{i+1}\rangle-
\langle \vec S_{i+1} \cdot \vec S_{i+2}\rangle}
{\frac{1}{2}\left[ \langle \vec S_{i} \cdot \vec S_{i+1}\rangle+
\langle \vec S_{i+1} \cdot \vec S_{i+2}\rangle\right] } \,.
\label{eq:sdimo}
\end{equation}

In Fig.\ \ref{fig:dim} we present our results for the short-range
dimer order parameter $S_{\rm srdo}$ measured in the middle of the
ladder as a function of the inverse of the chain length for a few
points of the  $(\lambda_1, \lambda_2=\lambda_3)$ phase space. The
parameter $R_{\rm srdo}$  not shown in the figure gives the same kind
of behavior.  

There is clearly an extended region in the parameter space, where the 
short-range order parameter scales to zero in the thermodynamic limit. 
At the VBS point itself $S_{\rm srdo}$ was found to be zero for short
finite  chains already. The vanishing of $S_{\rm srdo}$ happens not
only in the  Haldane phase, but also along the negative $\lambda_1$
axis and in a  neighborhood of it, where the spectrum was
undistinguishable from the  spectrum of the supposedly dimerized
state.  

For large negative values of $\lambda_2$, however, both for positive
and negative values of $\lambda_1$, a transition was found to the
state where  $R_{\rm srdo}$ and $S_{\rm srdo}$ have non-zero value
and the local  magnetization at the ends of the chain vanishes. Thus
in the region where  the ground state was found to be a
non-degenerate singlet, the short-range dimer order allows to
distinguish two regimes. In the region denoted  as `dimerized phase'
in Fig.\ \ref{fig:phase}, both $R_{\rm srdo}$ and $S_{\rm srdo}$
scale to a finite value. In this phase the local magnetization at the
ends of the chain is also absent. In the other region marked as
`massive', the dimer order disappears. The boundary between them is
drawn schematically only in Fig.\ \ref{fig:phase}, its exact shape
could  not be determined, except that it has to go through the $
\lambda_1 =  \lambda_2 = \lambda_3 =0$ point.

\section{Conclusions}

In the present paper we have considered a two-leg ladder model
constructed from a composite-spin model. Beside the usual Heisenberg
coupling between the spins on the legs inter-leg coupling between
spins on the same and neighboring rungs have been introduced, as well
as four-spin plaquette couplings. 

For special values of these couplings this ladder model is equivalent
to the spin-1/2 Heisenberg model or the spin-1 bilinear-biquadratic
model in the sense that the low-lying parts of the spectra are
identical. Thus  massless, VBS-like and dimerized phases are expected
to appear, but the  phase diagram can be even richer.

The behavior of the energy differences between low-lying levels has been 
calculated using the DMRG method and the gap extrapolated to the
thermodynamic limit ($N\to \infty$) was obtained with the method
of finite-size scaling. We have also considered the short-range dimer
order parameter. 

It has been found that the spin-1/2 Heisenberg ladder is unstable
against  most of the perturbations that couple the two legs, but
there is a small range of the parameters where the combination of the
perturbing operators is irrelevant. 

Four different kinds of behavior was observed, as shown in Fig.\  
\ref{fig:phase}. In a region along and near to the line $\lambda_1
=0$,  $\lambda_2 = \lambda_3 > 0$ the spectrum remains gapless.
Everywhere else a  gap is developed in the spectrum. Three regions
can, however, be  distinguished depending on whether the ground state
is non-degenerate,  twofold or fourfold degenerate. The last case is
easily detected by studying  the asymptotic degeneracy of the
spectrum. This phase is present around  the $\lambda_1 =1$,
$\lambda_2 = \lambda_3 =4/5$ point, where the exact  nearest-neighbor
valence-bond state is recovered. The extra $s=1/2$ spins  at the ends
of the chains become more and more delocalized as we move away  from
the VBS point and disappear at the phase boundary, where the
transition is either to the gapless phase or to a dimerized one.  

The Takhtajan-Babujian point at $\lambda_1=1$, $\lambda_2=\lambda_3 =
-4/3$  lies on the phase boundary between the VBS-like and the
dimerized phases. The transition here could be detected not only in
the change of the character of the spectrum, but also by the
appearance of a finite short-range dimer order parameter. This
quantity was used to distinguish the dimer phase  also from the
`massive' phase with non-degenerate singlet ground state,  where the
dimer order disappears again, in fact much faster than $1/N$.  

Unfortunately, our limited computational resources did not allow us to
determine the precise location of the critical lines separating the 
various phases.

\acknowledgments

This research was supported in part by the Hungarian Research Fund
(OTKA) Grant No.\ 15870, by the Swiss National Science Foundation
Grant No.\ 20-37642.93, the ISI Foundation (Torino) and the EU PECO
Network ERBCIPDCT940027.

\newpage

\begin{figure}
\caption{Schematic plot of the spin couplings in ${\cal H}_0$, ${\cal
H}_1$, ${\cal H}_2$ and ${\cal H}_3$ between the spins $\vec
\sigma_i$ and  $\vec \tau_i$ on the two legs of a ladder.}
\label{fig:ham}
\end{figure}

\begin{figure}
\caption{Low-lying energy spectrum of the Hamiltonian as a function of 
$\lambda_1$ $(\lambda_2 = \lambda_3 = 0)$ for chain length $N=6$.}
\label{fig:eigv}
\end{figure}

\begin{figure}
\caption{The energy difference between the low-lying levels of our
model  as a function of $\lambda_1$ at $\lambda_2= \lambda_3 =0 $ for
chain  lengths $8\leq N \leq 64$. For $\lambda_1 > 0$ $\Delta
E_{21}$, while for $\lambda_1 < 0$ $\Delta E_{10}$ is shown. The
dotted line is the gap  extrapolated for $N\rightarrow\infty$. The
inset shows the behavior for small values of $\lambda_1$.}
\label{fig:l1}
\end{figure}

\begin{figure}
\caption{The energy difference $\Delta E_{10}$ as a function of
$\lambda_2$ at $\lambda_1=\lambda_3 =0$ . The symbol x shows the
extrapolated value of  the gap for $\lambda_2 = -4/3$.}
\label{fig:l2}
\end{figure}

\begin{figure}
\caption{The quintuplet-triplet energy difference $\Delta E_{21}(N)$  
as a function of $\lambda_2= \lambda_3$ at $\lambda_1=1$. Close to the 
critical point the size of the symbol for the extrapolated gap
indicates  the error.}
\label{fig:l1_1a}
\end{figure}

\begin{figure}
\caption{The same as Fig.\ 5 for the triplet-singlet energy 
difference $\Delta E_{10}(N)$.}
\label{fig:l1_1}
\end{figure}

\begin{figure}
\caption{Schematic phase diagram of the model in the $(\lambda_1,
\lambda_2 = \lambda_3)$ plane. The wavy lines between the four
different phases indicates that the location of the phase boundaries
has not been determined accurately.} 
\label{fig:phase}
\end{figure}

\begin{figure}  
\caption{ The short-range dimer order parameter $S_{\rm srdo}$
calculated in the middle of the chain versus the inverse of the chain
length at a few points of the phase space $(\lambda_1, \lambda_2 =
\lambda_3)$.} 
\label{fig:dim}
\end{figure}


\begin{references}
\bibitem[*]{byline}
      {On leave from the Research Institute for Solid State Physics,
         Budapest, Hungary.}

\bibitem{johnston}{D.\ C.\ Johnston, J.\ W.\ Johnson, D.\ P.\ Goshorn,
    and A.\ J.\ Jacobson, Phys.\ Rev.\ B {\bf 35}, 219 (1987).}

\bibitem{hiroi}{Z.\ Hiroi, M.\ Azuma, M.\ Takano, and Y.\ Bando, J.\
    Solid State Chem.\ {\bf 95}, 230 (1991).}

\bibitem{cava}{R.\ J.\ Cava {\sl et al.}, J.\ Solid State Chem.\
      {\bf 94}, 170 (1991).}

\bibitem{haldane}{F.\ D.\ M.\ Haldane,
         Phys.\ Rev.\ Lett. {\bf 50}, 1153 (1983);
         Phys.\ Lett.\ {\bf 93A}, 464 (1983).}

\bibitem{hida}{K.\ Hida,  J.\ Phys.\ Soc.\ Jpn.\ {\bf 60} 1347
        (1991); {\bf 60} 1939 (1991). } 

\bibitem{strong}{S.\ P.\ Strong and A.\ J.\ Millis, Phys.\ Rev.\ Lett. 
        {\bf 69}, 2419 (1992).}

\bibitem{dagotto}{E.\ Dagotto, J.\ Riera, and D.\ Scalapino,
          Phys.\ Rev.\ B {\bf 45}, 5744 (1992).}

\bibitem{watanabe}{S.\ Takada and H. Watanabe, J.\ Phys.\ Soc.\ Jpn.\
    {\bf 61}, 39 (1992); H.\ Watanabe, K.\ Nomura, and S.\ Takada,
         {\rm ibid}.\ {\bf 62} 2845 (1993); H.\ Watanabe,
         Phys.\ Rev.\ B {\bf 50}, 13442 (1994).}

\bibitem{barnes}{T.\ Barnes, E.\ Dagotto J.\ Riera, and E.\ S.\ Swanson,
         Phys.\ Rev.\ B {\bf 47}, 3196 (1993); T.\ Barnes and J.\ Riera, 
         {\em ibid}.\ {\bf 50}, 6817 (1994).}

\bibitem{hsu}{T.\ Hsu and J.\ C.\ Angl\`es d'Auriac, 
          Phys.\ Rev.\ B {\bf 47}, 14291 (1993).}

\bibitem{noack}{R.\ M.\ Noack, S.\ R.\ White, and D.\ J.\ Scalapino,
         Phys.\ Rev.\ Lett. {\bf 73}, 882 (1994);
        S.\ R.\ White, R.\ M.\ Noack, and D.\ J.\ Scalapino,
         {\em ibid}. {\bf 73}, 886 (1994).}

\bibitem{gopalan}{S.\ Gopalan, T.\ M.\ Rice, and M.\ Sigrist,
    Phys.\ Rev.\ B {\bf 49}, 8901 (1994); M.\ Sigrist, T.\ M.\ Rice, 
         and F.\ C.\ Zhang, {\em ibid}. {\bf 49}, 12058 (1994).}

\bibitem{tsune}{H.\ Tsunetsugu, M.\ Troyer, and T.\ M.\ Rice,
   Phys.\ Rev.\ B {\bf 49}, 16078 (1994); M.\ Troyer, H.\ Tsunetsugu, 
         and D.\ W\"urtz, {\em ibid}. {\bf 50}, 13515 (1994).}

\bibitem{totsu}{K.\ Totsuka and M.\ Suzuki,
        J.\ Phys.: Condens. Matter {\bf 7}, 6079 (1995).}

\bibitem{white95}{S.\ R.\ White, Phys.\ Rev.\ B {\bf 53}, 52 (1996).}

\bibitem{matter1}{R.\ S.\ Eccleston, T.\ Barnes, J.\ Brody, and J.\ W.\
Johnson, Phys.\ Rev.\ Lett.\ {\bf 73}, 2626 (1994).}

\bibitem{azuma}{M.\ Azuma, Z.\ Hiroi, M.\ Takano, K.\ Ishida, and Y.\
Kitaoka, Phys.\ Rev.\ Lett.\ {\bf 73}, 3463 (1994).}

\bibitem{solyom1}{J.\ S\'olyom and J.\ Timonen,
         Phys.\ Rev.\ B {\bf 34}, 487 (1986); {\bf 38}, 6832 (1988); 
       {\bf 39}, 7003 (1989).}

\bibitem{schulz}{H.\ J.\ Schulz, Phys.\ Rev.\ B {\bf 34}, 6372 (1986).}

\bibitem{white}{S.\ R.\ White,
         Phys.\ Rev.\ Lett.\ {\bf 69}, 2863 (1992);
         Phys.\ Rev.\ B {\bf 48}, 10345 (1993).}

\bibitem{aff}{I.\ Affleck, Phys.\ Rev.\ Lett. {\bf 55}, 1355 (1985);
         Nucl.\ Phys.\ B {\bf 265}, 409 (1986)}

\bibitem{takt}{L.\ Takhtajan, Phys.\ Lett. {\bf 87A}, 479 (1982);
         H.\ M.\ Babujian, {\sl ibid}. {\bf 90A}, 479 (1982)}

\bibitem{lai}{C.\ K.\ Lai, J.\ Math.\ Phys. {\bf 15}, 1675 (1974);
          B.\ Sutherland, Phys.\ Rev.\ B {\bf 12}, 3795 (1975). }

\bibitem{vbs}{I.\ Affleck, T.\ Kennedy, E.\ H.\ Lieb, and H.\ Tasaki,
         Phys.\ Rev.\ Lett. {\bf 59}, 799 (1987);
         Commun.\ Math.\ Phys. {\bf 115}, 477 (1988).}
          
\bibitem{parkinson}{J.\ B.\ Parkinson, J.\ Phys.\ C {\bf 21}, 
     3793 (1988); M.\ N.\ Barber and M.\ T.\ Bachelor, 
      Phys.\ Rev.\ B {\bf 40}, 4621 (1989):
   A.\ Kl\"umper, J.\ Phys.\ A {\bf 23}, 809 (1990).}

\bibitem{kung}{D.\ Kung (unpublished); J.\ Oitmaa, J.\ B.\ 
    Parkinson, and  J.\ C.\ Bonner, J.\ Phys.\ C {\bf 19}, L595
    (1986); H.\ W.\ J.\ Bl\"ote  and H.\ W.\ Capel, Physica {\bf
     139A}, 387 (1986).} 

\bibitem{solyom}{J.\ S\'olyom, Phys.\ Rev.\ B {\bf 36}, 8642 (1987).}

\bibitem{tak}{K.\ Nomura and S.\ Takada,
           J.\ Phys.\ Soc.\ Jpn. {\bf 60}, 389 (1991).}

\bibitem{gfath}{G.\ F\'ath and J.\ S\'olyom,
         Phys.\ Rev.\ B {\bf 44}, 11836 (1991);
         {\em ibid.} {\bf 47}, 872 (1993).}

\bibitem{itoi}{C.\ Itoi and M.\ H.\ Kato, report No. cond-mat/9605105 
            (unpublished).}

\bibitem{chubukov}{A.\ V.\ Chubukov, Phys.\ Rev.\ B {\bf 43}, 
       3337 (1991). }

\bibitem{gfath95}{G.\ F\'ath and J.\ S\'olyom,
         Phys.\ Rev.\ B {\bf 51}, 3620 (1995).}

\bibitem{legeza}{\"O.\ Legeza and G.\ F\'ath, 
         Phys.\ Rev.\ B {\bf 53}, 14349 (1996).}

\bibitem{sorensen}{E.\ S.\ Sorensen and I.\ Affleck,
         Phys.\ Rev.\ Lett. {\bf 71}, 1633 (1993).}

\bibitem{schollwock}{U.\ Schollw\"ock and T.\ Jolicoeur,
                         Europhys.\ Lett.\ {\bf 30}, 493 (1995).}

\bibitem{white3}{S.\ R.\ White and D.\ A.\ Huse
         Phys.\ Rev.\ B. {\bf 48}, 3844 (1993).}

\bibitem{gfath3}{G.\ F\'ath and J.\ S\'olyom,
         J.\ Phys.: Condens. Matter {\bf 5}, 8983 (1993).}

\end{references}
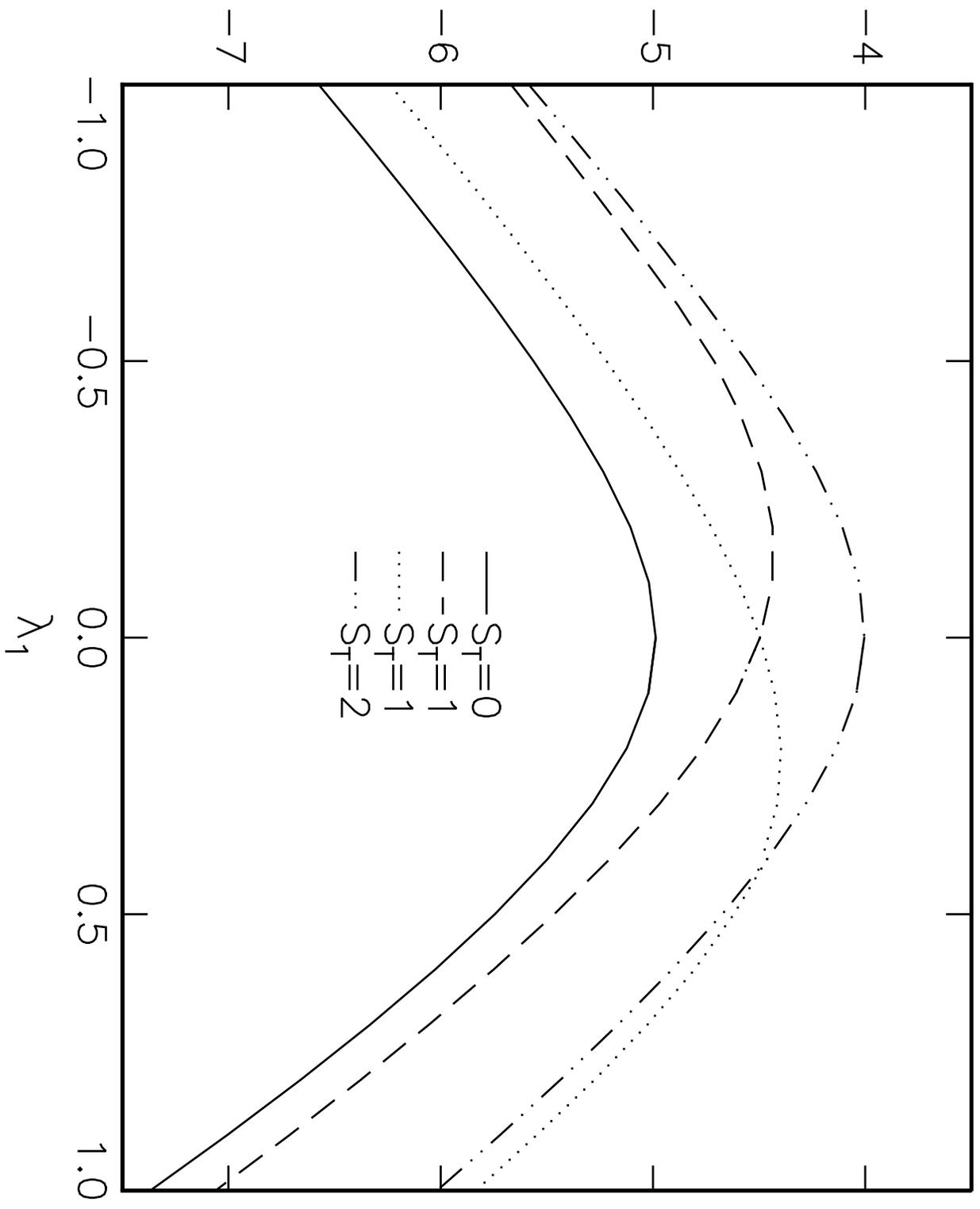
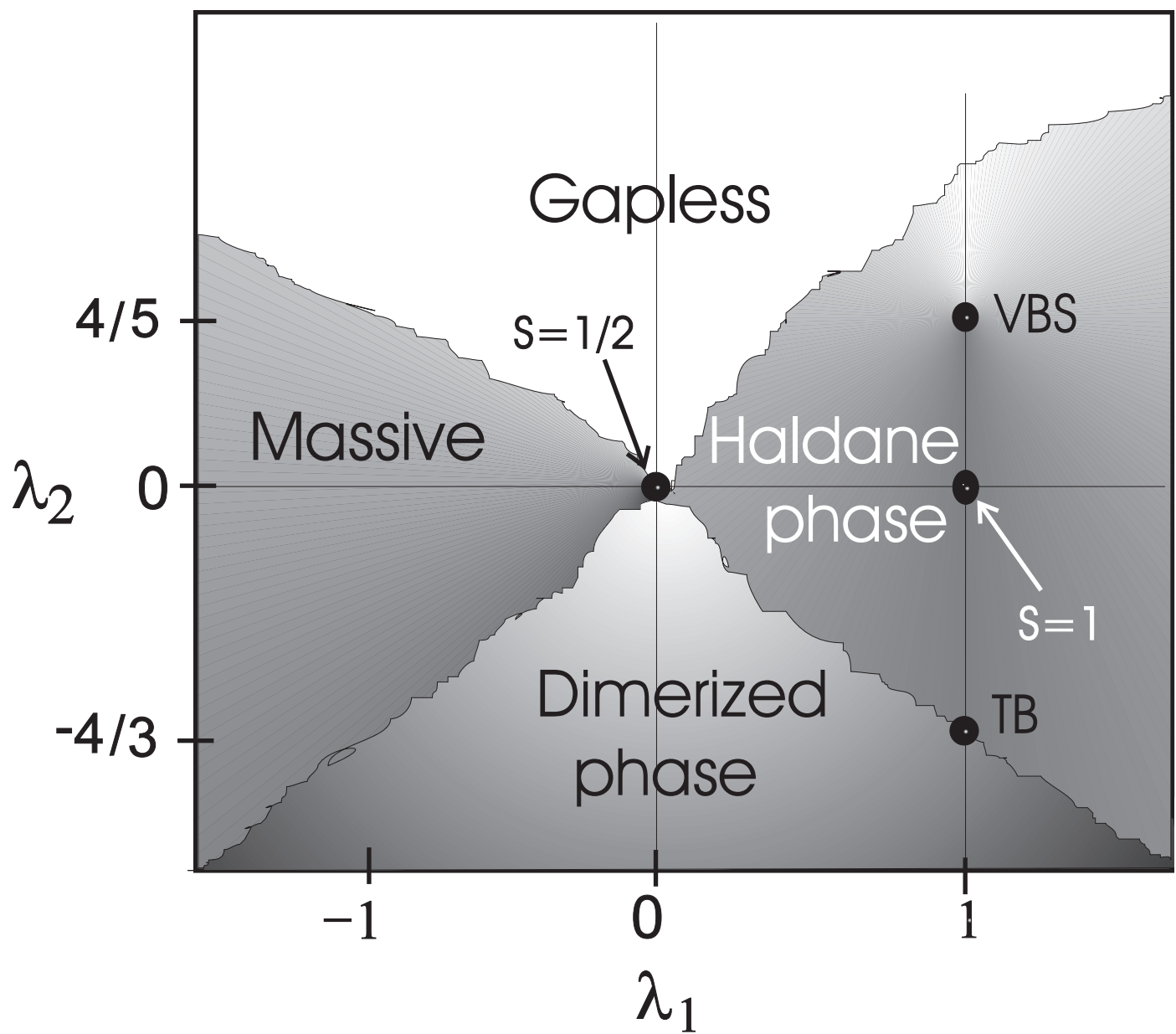
\end{document}